\documentclass[twocolumn,prl,showpacs]{revtex4}
\usepackage{graphicx}
\usepackage{dcolumn}
\usepackage{amsmath}
\usepackage{bm}
\newcommand{\be}{\begin{equation}}
\newcommand{\ee}{\end{equation}}
\newcommand{\ba}{\begin{eqnarray}}
\newcommand{\ea}{\end{eqnarray}}
\begin{document}
\title{Hexatic phase in the two-dimensional Gaussian-core model}
\author{Santi Prestipino$^1$~\cite{aff1}, Franz Saija$^2$~\cite{aff2},
and Paolo V. Giaquinta$^1$~\cite{aff3}}
\affiliation{
$^1$ Universit\`a degli Studi di Messina, Dipartimento di Fisica,
Contrada Papardo, I-98166 Messina, Italy \\
$^2$ CNR-IPCF, Viale Ferdinando Stagno d'Alcontres 37, 98158 Messina, Italy}
\date{\today}
\begin{abstract}
We present a Monte Carlo simulation study of the phase behavior of
two-dimensional classical particles repelling each other through an
isotropic Gaussian potential.
As in the analogous three-dimensional case, a reentrant-melting transition
occurs upon compression for not too high temperatures, along with a
spectrum of water-like anomalies in the fluid phase. However, in two
dimensions melting is a continuous two-stage transition, with an intermediate
hexatic phase which becomes increasingly more definite as pressure grows.
All available evidence supports the Kosterlitz-Thouless-Halperin-Nelson-Young
scenario for this melting transition.
We expect that such a phenomenology can be checked in confined
monolayers of charge-stabilized colloids with a softened core.

\end{abstract}
\maketitle

In two dimensions thermal fluctuations do not allow the existence of a true
crystalline order; in fact, only a quasi-long-range translational order is
possible while bond-angular order is truly long-ranged. This opens the way
to a two-stage melting transition through an intermediate ``hexatic'' phase
with short-ranged translational order but extended bond-angle correlations.
In the celebrated Kosterlitz-Thouless-Halperin-Nelson-Young (KTHNY) theory
of two-dimensional (2d) melting~\cite{KTHNY}, the hexatic phase is promoted
by the thermal unbinding of dislocation pairs, followed by the proliferation of
free disclinations on entering the normal fluid. The KTHNY theory predicts
melting to be continuous.
In two dimensions, when the energy of the dislocation core is sufficiently
small, a first-order melting transition is more likely, driven by the
spontaneous generation of grain boundaries~\cite{Chui,Strandburg}.
Hexatic phases have been observed in various types of
colloids~\cite{Murray1,Marcus,Zahn,Dullens,Vongruenberg,Keim,Lin1,Han,Peng},
and found also in some classical~\cite{Jaster,Muto,Lin2,Lee,Qi} and quantum
simulations~\cite{Clark}.
Moreover, nothing prevents the hexatic phase to be just metastable,
as observed e.g. in Ref.\,\cite{Chen}.

Observing the KTHNY scenario is notoriously difficult because of the
existence of important finite-size effects and long equilibration times.
Also, the usually narrow temperature extent of the hexatic phase makes
it hard to distinguish a two-stage melting from a single weakly first-order
transition. Particularly severe is the situation for hard-core particles,
where enormous samples and huge simulation times are required in order to
discriminate between the various transition scenarios~\cite{Bernard}, while
less demanding may be state sampling for systems of ``soft'' particles
whose steric constraints are less pronounced.

\begin{figure}
\includegraphics[width=8cm]{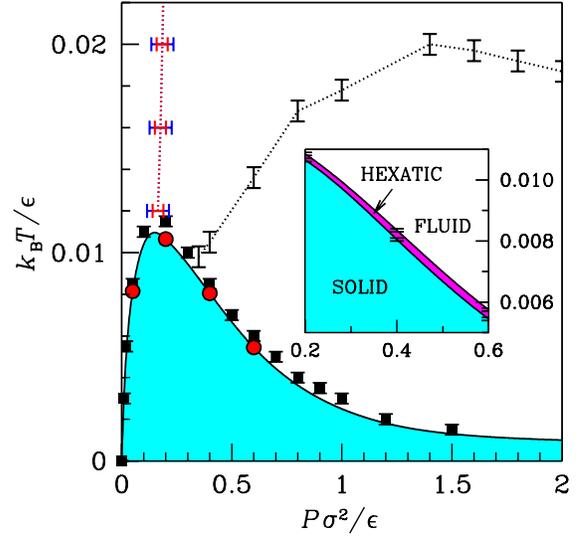}
\caption{(Color online). Phase diagram of the 2d GCM with anomaly loci.
Plotted in black is the melting line, with red dots at the computed
solid-to-hexatic transition points. The black squares (with error bars
superimposed) give the upper stability threshold of the solid when
heated isobarically in steps of $\Delta T=0.0005$.
The inset shows a magnified portion of the melting line with an adjacent
(magenta) strip corresponding to the hexatic region. The dotted curves
mark the boundary of anomaly regions (structural anomaly, red;
diffusion anomaly, blue; density anomaly, black). Observe that the red and
blue curves, which appear indistinguishable on the scale of the figure,
depart from each other at much higher temperatures.}
\label{fig1}
\end{figure}

We hereby inquire into the existence of a hexatic phase for the 2d
Gaussian-core model (GCM) pair potential~\cite{Stillinger1},
$v(r)=\epsilon\exp(-r^2/\sigma^2)$ with $\epsilon>0$, which is somewhat
representative of a whole class of systems of interpenetrating particles
(e.g. dilute dispersions of polymer chains)~\cite{Likos1}.
In three dimensions, this
system is known to exhibit reentrant melting ({\it i.e.}, melting upon
compression at constant temperature)~\cite{Prestipino} as well as waterlike
anomalies~\cite{Mausbach}.
Except for a 30-year old canonical-ensemble investigation~\cite{Stillinger2}
with inconclusive answers, we do not know of
any simulation study of the melting behavior of 2d systems of particles
with bounded interactions with a focus on the quest for a hexatic phase.
Furthermore, it would be interesting to know about the interplay between
anomalous melting (that is, melting of the solid into an anomalous fluid)
and the modality of decay of bond-angle correlations, an issue that has never
been addressed before. As discussed in more detail below, the melting of
the 2d GCM is indeed continuous and two-staged, with an extremely narrow
hexatic region whose properties comply with the
predictions of the KTHNY theory. The complete phase diagram is plotted
in Fig.\,1, together with a number of anomaly loci in the fluid phase.

Particles interacting through a repulsive Gaussian potential are expected
to exhibit reentrant melting and a maximum melting temperature~\cite{Likos2}.
By examining all the five Bravais lattices and the honeycomb lattice, we first
checked that the most stable state of the GCM at zero temperature is a
triangular crystal for any pressure $P$.
This gave us confidence that the triangular lattice provides the structure
of the solid phase also for non-zero temperatures. We carried out
isothermal-isobaric Monte Carlo (MC) simulations of $N$-particle samples
(with $N$ up to 6048) in order to locate melting for a number of selected
pressures (0.05, 0.2, 0.4, and 0.6, in reduced, $\epsilon/\sigma^2$
units). Our method consists in running simulations in a sequence, starting
from the cold triangular solid on one side of the chain and from the hot fluid
on the other side. Then, the solid was gradually heated (the fluid was cooled)
in temperature steps of $\Delta T=0.0005$ (in reduced, $\epsilon/k_B$ units),
until we observed the abrupt melting (freezing) of the system.
With this protocol, we found the same shape of the melting line as in the
three-dimensional GCM, with a maximum melting temperature $T_m$ of
about 0.0115 for $P=P_m\lesssim 0.2$. We plot in Fig.\,1 the melting line
with three other curves which encompass regions in the fluid phase where an
``anomalous'' behavior occurs.
On increasing the density, one first meets the so-called structural
anomaly (which is where the absolute pair entropy reaches its
maximum)~\cite{Giaquinta},
followed by the diffusion-anomaly locus (where the self-diffusion
coefficient~\cite{note} attains its minimum), and by the density-anomaly line
(where the particle-number density attains a local maximum). The same
succession of anomaly loci is found in three dimensions~\cite{Krekelberg}.

To disentangle first-order from continuous melting, we performed another
series of runs across our earlier guess of the
transition point, now with a 5-time larger $T$ resolution and also
allowing for much longer equilibration times ($10^6$ sweeps, that is
a million MC moves per particle) and production runs ($5\times 10^6$
sweeps). A typical result is shown in Fig.\,2, where we report the
average specific energy $u$ and particle-number density $\rho$ for various
system sizes as a function of $T$ for $P=0.6$. A continuous path joins
the solid and fluid branches with no evidence of hysteresis,
which points to a smooth transition between the solid and fluid phases.
Moreover, the energy and volume histograms have a simple Gaussian shape
with no trace of bimodality within the relevant temperature range.
As we are going to show in the following, the intermediate region
between the solid and the (normal) fluid can be qualified as hexatic.

\begin{figure}
\includegraphics[width=8cm]{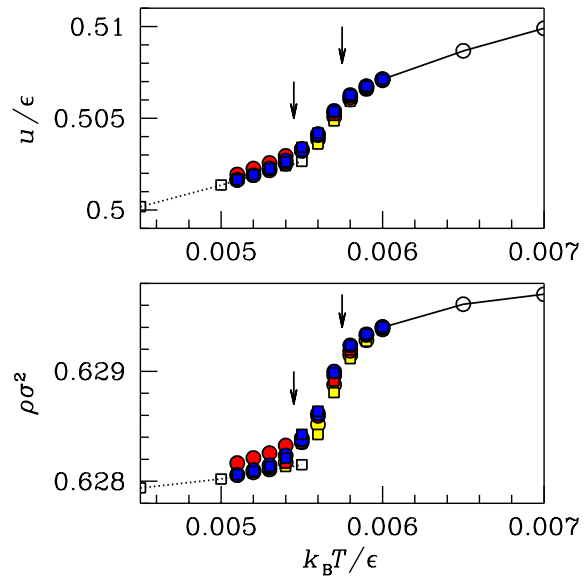}
\caption{(Color online). Total energy per particle (top) and particle-number
density (bottom) for three different sizes ($N=1152$, yellow; $N=2688$,
red; $N=6048$, blue) for $P=0.6$. We show results for both heating (squares)
and cooling (dots) trajectories (one million MC sweeps of equilibration
plus five million sweeps of data accumulation). The solid and fluid
branches (dotted and solid lines, respectively) are computed with
$N=1152$ and much smaller statistics. The arrows mark the estimated
transition points (see Fig.\,3). The small hysteresis observed upon cooling
for $N=2688$ indicates that much longer runs are needed in order that the
solidifying system may get rid of the extra defects.}
\label{fig2}
\end{figure}

We measured two different order parameters (OP), which are separately
sensitive to the overall translational and orientational triangular order,
with their respective susceptibilities and correlation functions.
The translational OP is taken to be
\be
\psi_T=\frac{1}{N}\left\langle\left|\sum_ie^{{\rm i}{\bf G}\cdot{\bf r}_i}
\right|\right\rangle\,,
\label{eq1}
\ee
where the sum is over the particle labels and {\bf G} is any first-shell
reciprocal-lattice vector of the triangular crystal. From its very definition,
it follows that $\psi_T$ is sizeable only in a triangular solid
that is oriented in a way consistent with the length {\it and} direction
of {\bf G}. Hence, $\psi_T$ is only measured on heating, where memory of
the original crystal orientation is preserved as long as the system is large
and remains solid. We anyway checked -- through the location of the main peaks
of the structure factor -- that the orientation of the solid never changed
from one run to the next.
A sharp drop of $\psi_T$ signals the melting of the solid into a fluid,
be it hexatic or normal; concurrently, the corresponding susceptibility
\be
\chi_T=\frac{1}{N}\left\langle\left|\sum_ie^{{\rm i}{\bf G}\cdot{\bf r}_i}
\right|^2\right\rangle-N\psi_T^2\,,
\label{eq2}
\ee
shows a distinct peak whose location is an unambiguous estimate
of the melting transition point. At regular intervals during the simulation,
we made use of the Voronoi construction in order to identify the $n_c(i)$
nearest neighbors (NN) of each particle $i$, together with the orientation
$\theta_{NN}$ of each neighbor bond with respect to a reference axis.
Whence, the orientational OP follows as
\be
\psi_6=\frac{1}{N}\left\langle\left|\sum_i\frac{1}{n_c(i)}
\sum_{NN(i)}e^{6{\rm i}\theta_{NN}}\right|\right\rangle\equiv
\frac{1}{N}\left\langle\left|\sum_i\Psi_6({\bf r}_i)\right|\right\rangle\,.
\label{eq3}
\ee
The orientational susceptibility $\chi_6$ is then defined in a way
analogous to Eq.\,(\ref{eq2}), with $\Psi_6({\bf r}_i)$ replacing
$\exp({\rm i}{\bf G}\cdot{\bf r}_i)$. $\psi_6$ undergoes a sudden drop
at the hexatic-fluid transition, {\it i.e.}, at a temperature
larger than the one where $\psi_T$ vanishes.
Finally, the local bond-angular OP $\Psi_6({\bf r}_i)$ enters the
definition of the orientational correlation function (OCF):
\be
h_6(r)=\rho^{-2}\left\langle\sum'_{i,j}\delta^3({\bf r}_i-{\bf R})
\delta^3({\bf r}_j-{\bf R}')\Psi_6({\bf r}_i)\Psi_6^*({\bf r}_j)
\right\rangle\,,
\label{eq4}
\ee
where the prime over the sum excludes $i=j$ and $r=|{\bf R}-{\bf R}'|$.
The KTHNY theory predicts an algebraic $r^{-\eta(T)}$ large-distance
decay of the OCF in the hexatic phase, which should be contrasted with the
exponential asymptotic vanishing of angular correlations in a normal fluid.
Another prediction of the theory is $\eta=1/4$ at the hexatic-to-normal
fluid transition point.

\begin{figure}
\includegraphics[width=8cm]{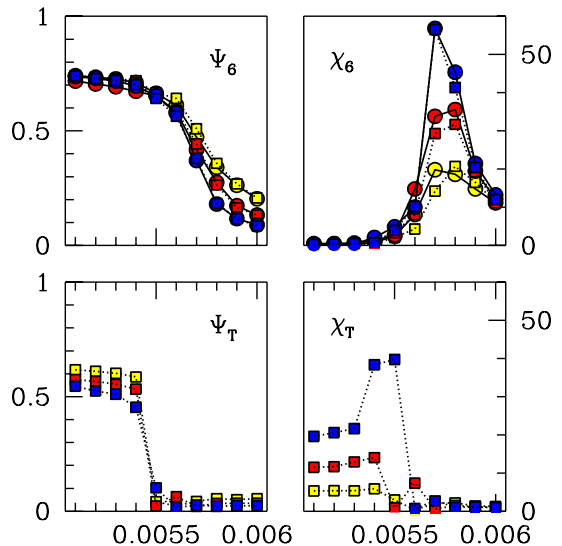}
\caption{(Color online). Order parameters and susceptibilities for
$P=0.6$ in the $T$ range across the melting transition. Upper panels:
the orientational order parameter $\psi_6$ and its susceptibility $\chi_6$
for three system sizes (color codes as in Fig.\,2).
Dots and squares mark data obtained by cooling and by heating, respectively.
Roughly, the difference between the two estimates gives a clue about the
statistical uncertainty associated with each data point. Lower panels:
the translational order parameter $\psi_T$ and its susceptibility $\chi_T$
for the same sizes on heating. The non-zero value of $\psi_T$ in the solid
phase is actually a finite-size effect, made possible by the use of periodic
boundary conditions in the simulations, since for a 2d {\em infinite} solid
quasi-long-range translational order implies $\psi_T=0$. Moreover, $\chi_T$
is expected to diverge in the solid phase of an infinite-size system. Similar
considerations apply for the behavior of $\psi_6$ and $\chi_6$ in the
hexatic phase.}
\label{fig3}
\end{figure}

In Fig.\,3, we plot the two OPs and susceptibilities for $P=0.6$ (an
analogous behavior was observed for all the other pressures).
We see that $\psi_T$ vanishes at a slightly smaller temperature than
$\psi_6$, which implies that the hexatic phase is confined to an extremely
narrow $T$ interval not wider than 0.0002--0.0003, as also witnessed
by the maxima of the two susceptibilities occurring at slightly different
$T$ values. The estimated width of the hexatic region compares well with
the temperature range of the bridging region between the solid and
fluid branches in Fig.\,2. While the size scaling of $\chi_6$ is a clear
imprint of a second-order hexatic-to-normal fluid transition, the
solid-to-hexatic transition might even be first-order, were this not in
contrast to the smooth behavior of $u$ and $\rho$. Upon reducing the pressure,
the width of the hexatic phase gradually shrinks until, for $P=0.05$,
it becomes comparable to the temperature resolution. However, even
in this case we tend to exclude the disappearance of the hexatic phase for
low pressures since this would imply the existence of a triple point for
which we currently have no independent evidence.
Finally, it is worth mentioning the case $P=0.2$ (a pressure above
$P_m$ but outside the density-anomaly region), where the hexatic
fluid shows a density anomaly while the normal fluid does not --
in no other way could the density branch of the normal fluid have hooked
on a solid branch that lies at a lower density level.

A more direct evidence of the hexatic phase emerges from the large-distance
behavior of the OCF. We plot this function in Fig.\,4 at various
temperatures across the hexatic phase for $P=0.6$.
It appears that the OCF decays algebraically in a $T$ region of
limited extent, which roughly corresponds to the middle of the bridging
region in Fig.\,2. Moreover, the decay exponent in this hexatic region is
smaller than $1/4$, becoming larger only on passing to the normal fluid.

\begin{figure}
\includegraphics[width=8cm]{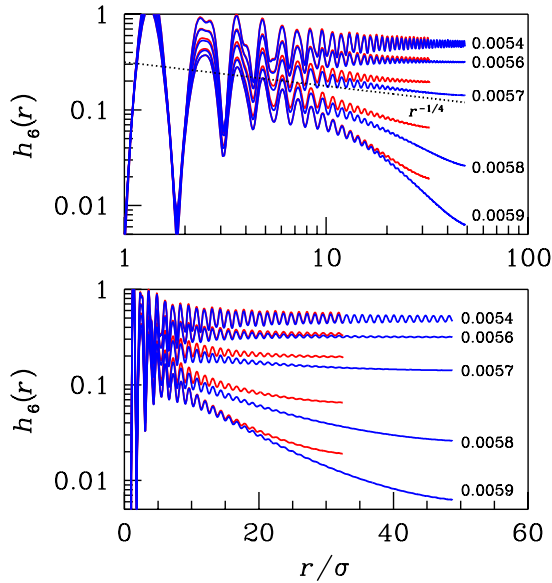}
\caption{(Color online). Orietational correlation function $h_6(r)$ at
selected temperatures across the hexatic
region for $P=0.6$. We plot $h_6(r)$ on heating for two sizes, $N=2688$
(red) and $N=6048$ (blue). Top: log-log plot; bottom: log-lin plot.
Upon increasing $T$ from 0.0054 to 0.0059 there is a qualitative
change in the large-distance behavior of $h_6(r)$, from constant (solid)
to power-law decay (hexatic fluid), up to exponential decay (normal fluid).
Note that, consistently with the KTHNY theory, the decay exponent $\eta$
is less than 1/4 ({\it i.e.}, the slope of the dotted curve) in the hexatic
phase.}
\label{fig4}
\end{figure}

We finally checked a further KTHNY prediction concerning
the behavior of a 2d triangular solid which is about to melt into a hexatic
phase. The elastic instability that signals the onset of dissociation of
dislocation pairs, preluding to the stabilization of the hexatic phase, is
heralded by the value of
\be
K=\frac{4a^2}{k_BT}\frac{\mu(\mu+\lambda)}{2\mu+\lambda}
\label{eq5}
\ee
becoming equal to $16\pi$~\cite{KTHNY,Bates}.
In Eq.\,(\ref{eq5}), $\lambda$ and $\mu$ are the Lam\'e coefficients (as
renormalized by the thermal fluctuations) while $a=\sqrt{2/(\sqrt{3}\rho)}$
is the lattice parameter.
$\lambda$ and $\mu$ are respectively given by $c_{12}+P$ and $c_{44}-P$,
in terms of the elastic constants $c_{12}$ and $c_{44}$ which can be computed
as canonical-ensemble averages from virial-like formulae~\cite{Farago}.
We found an impressive confirmation of the theory for $P=0.6$ while the
threshold value of $K/(16\pi)$ (before its drop to zero) turned out to be
a bit larger than one (1.1--1.2) for the other investigated pressures.
This further indicates that the overall KTHNY picture
deteriorates with reducing pressure,
probably because the formation energy of a dislocation becomes
smaller and smaller with increasing average interparticle distances.

In conclusion, we have provided the first unambiguous evidence of the
occurrence of two-stage continuous reentrant melting via a hexatic phase
in the 2d Gaussian-core model, taken as prototypical of the
phase behavior of bounded model potentials.
We have validated a number of KTHNY predictions,
though larger samples and more statistics will be
necessary in order to ascertain the real nature of melting
at low densities. The present discovery of reentrant-hexatic behavior in
the GCM is relevant for many soft-matter systems. For instance, one can
engineer colloidal particles interacting through a temperature-modulated
softened repulsion, which will likely exhibit GCM-like reentrant melting
in a range of packing fractions well below the density at which hard-core
crystallization occurs (see \cite{Royall} for a 3d realization of this
scenario). Such systems would be natural candidates where to detect
(e.g. by video microscopy~\cite{Murray2}) a reentrant-hexatic phenomenon
of the kind illustrated here.


\begin{thebibliography}{99}
\bibitem{note} The self-diffusion coefficient $D$ was computed for
$N=1152$, via molecular dynamics, from the long-time behavior of the mean
square displacement (MSD) of the particles; we found that the MSD increases
linearly with time over a wide time interval and does not show any
significant size dependence in the region explored.
\end{thebibliography}
\end{document}